\documentclass[preprint]{revtex4}
\usepackage{amsmath}
\usepackage{color}
\usepackage{amssymb}
\usepackage{graphicx}
\usepackage[utf8]{inputenc}
\usepackage{textcomp}
\usepackage{color}
\usepackage{esint}

\begin{document}

\title{Effective Mass and Feld-Reinforced Superconductivity in Uranium Compounds}

\author{Vladimir P. Mineev}

\affiliation{Landau Institute for Theoretical Physics, 142432 Chernogolovka, Russia}

\begin{abstract}

A theory of strong coupling superconductivity in uranium compounds has been developed, based on electron-electron interaction through magnetic fluctuations described by frequency-dependent magnetic susceptibility. The magnetic field dependence of the electron effective mass is expressed through the field dependence of the magnetic susceptibility components.
   It is shown that the intensity of triplet pairing, and hence the critical temperature of the transition to the superconducting state, is also determined by the field-dependent susceptibility. The results are discussed in relation to the properties of ferromagnetic uranium compounds URhGe and UCoGe, as well as the recently discovered UTe$_2$.

\end{abstract}
\date{\today}
\maketitle

\section{Introduction}

The standard orbital mechanism suppressing superconducting state  is the depairing caused by magnetic field.
In addition the intensity of pairing  itself 
can  decrease or increase depending on the magnitude of field. The latter possibility  violates  the simple monotonic decrease in the critical temperature and can lead to a peculiar phenomenon of reentrant superconductivity. This situation is realised 
in uranium ferromagnetic superconductors URhGe, UCoGe   for the field direction parallel to the b-axis, perpendicular to the spontaneous magnetisation \cite{Flouquet2019}.
The former possibility is realised in UCoGe for field parallel to the spontaneous magnetisation and reveals itself as the upward curvature of temperature dependence of the upper critical field \cite{Wu2017}.

In the theory of strong coupling  superconductivity  the critical temperature \cite{McMillan1968}  $T_{c}=\omega_D\exp\left (-\frac{1+\lambda}{\lambda}\right)$ depends on the effective mass of the electron
$m^{\star}=m(1+\lambda)$ renormalised by electron-phonon interaction.
And if the effective mass turns out to be dependent on the magnetic field, then one can count on obtaining the field-dependent intensity of the pair interaction.
 This type of field dependence of pairing intensity has been 
proposed in the paper \cite{Miyake2008} and  later  used in many publications (see review \cite{Flouquet2019}). The dependence  $\lambda({\bf H})$
is extracted from the field dependence of specific heat and   of $A$ coefficient  in the low temperature resistivity behaviour $\rho(T)=\rho_0+AT^2$.
 However, it remained unclear why the physics of pairing interaction in uranium superconductors is described by a   theory that is valid for superconducting state with singlet pairing,  but with a field dependent parameter 
$\lambda({\bf H})$.

The theoretical explanations of the field dependence of pairing intensity have been proposed by the author \cite{Mineev2011,Mineev2017} and  by K.Hattori and H.Tsunetsugu \cite{Hattori2013}.
These approaches were based  not on the field dependence of the effective mass, but on the field dependence of the  interaction of electrons due to magnetisation fluctuations localised predominantly on uranium ions. Within the framework of this approach, it was also proven that the effective mass depends on the magnetic field, but not in the same way as the pairing amplitude \cite{Mineev2020}.

In URhGe, reentrant superconductivity occurs near the metamagnetic phase transition at a field $H_b\approx12.5$ T not far from the critical end point of the metamagnetic transition line. In the same field region, the $1/T_{2,b}$ NMR relaxation rate tends to diverge \cite{Tokunaga2015,Tokunaga2016}. Static susceptibility also diverges near the metamagnetic transition line \cite{Nakamura2017}
as it should be according to the theory of critical phenomena in the vicinity of the van der Waals critical end point
   \cite{Mineev2021}. The effective mass  increases with field parallel to b-axis \cite{Hardi2011}.

In UCoGe, stimulation of superconductivity in a field parallel to the $b$ axis is also associated with an increase in susceptibility. But in this case, the latter occurs due to the suppression of the Curie temperature by a magnetic field parallel to the b axis \cite{Ishida2021}. The effective mass in a field parallel to the $b$ axis increases \cite{Aoki2014}. On the contrary, a growth of magnetic field in the direction of spontaneous magnetisation causes a decrease in magnetic susceptibility and is accompanied by a decrease in the effective electron mass and the critical temperature of superconductivity \cite{Wu2017}.

 The uranium superconductor UTe$_2$, discovered about four years ago \cite{Ran-Science, Aoki2019}, has many unusual properties \cite{Aoki2022}. It is an orthorhombic paramagnetic metal with an easy magnetisation axis parallel to the $a$-crystallographic direction and a critical transition temperature to the superconducting state of about 2.0 K.
The most impressive observation\cite{Ran-NatPhys,Knebel2019} was that the superconducting state
   UTe$_2$ in a magnetic field oriented along the $b$ axis
   persists up to 34.5 T, where superconductivity is destroyed by the metamagnetic transition. Thermodynamic measurements indicate the formation of a new superconducting phase in fields above 15 Tesla \cite{Rosuel2023,Sakai2023}. Effective mass in a field parallel
  the $b$ axis increases strongly as the field approaches the metamagnetic transition \cite{Imajo2019,Miyake2019,Miyake2021}.
Recently published NMR measurements in UTe$_2$ in a field along the $b$-crystallographic axis carried out by
   Y. Tokunaga {\it et al } \cite{Tokunaga2023} demonstrate a strong increase in the intensity of longitudinal magnetic fluctuations in fields above 15 Tesla.
    This looks like a serious hint about the reason for the appearance of a reentrant superconducting state in UTe$_2$ in a strong magnetic field along the $b$-crystallographic direction.

 Thus, in all mentioned uranium compounds there are several phenomena, which, apparently, are somehow related to each other. These are a field dependence of  the effective mass of electrons, the appearance reentrant superconductivity, the field dependent growth  of NMR relaxation rates and magnetic susceptibility . To find a connection between these phenomena is the goal of present article.

The previous consideration of superconductivity in uranium compounds, developed by the author, was carried out within the framework of the weak coupling theory \cite{Mineev2011,Mineev2017}.
In this paper there will be developed a strong coupling theory of superconductivity with triplet pairing.
Similar  to the traditional theory of superconductivity with singlet pairing using retarded electron-phonon interaction here I will work with
electron-electron interaction generated by magnetic fluctuations described by frequency-dependent magnetic susceptibility.
 For the first time is derived
   a formula expressing the effective mass of an electron through the components of magnetic susceptibility and, thereby, establishing a connection between the field and temperature dependence of two independently measured quantities. This result also allows  to qualitatively interpret the behaviour of NMR relaxation rates.
  In neglect of orbital effects there was found the critical temperature of transition to superconducting state in a ferromagnetic metal with a conduction band split by exchange interaction into bands with spin up and spin down electrons. 
  The orbital effects are taken into account qualitatively by making use 
  the field dependence of  the Fermi velocity and 
   the critical temperature in neglect of orbital effects.

The paper is organised as follows. In the next section we will formulate the basic equations and obtain explicitly the effective mass magnetic field dependence through the field depending magnetic susceptibility. Then, in neglect of  orbital effects there will be derived a formula for the field dependent critical temperature of transition to superconducting state also expressed through magnetic susceptibility. The obtained results are compared with observed properties of URhGe, UCoGe and UTe$_2$. Finally, the qualitative treatment of orbital effects is presented allowing explain the phenomenon of the reentrant superconductivity.

  \section{Effective mass}

We consider the interaction between  the electrons by means self-induced magnetic polarisation
\begin{equation}
V(t)=-\frac{1}{2}g^2\int d^3{\bf r}d^3{\bf r}^\prime \int dt^\prime S_i({\bf r},t)\chi_{ij}({\bf r-{\bf r}^\prime},t-t^\prime)S_j({\bf r}^\prime,t^\prime).
\label{int}
\end{equation}
Here, 
$$
{\bf S}({\bf r},t)=\psi^\dagger_\alpha({\bf r},t)\mbox{\boldmath$\sigma$}_{\alpha\beta}\psi_\beta({\bf r},t)
$$
is the operator of the electron spin density, $\chi_{ij}({\bf r},t) $ is the magnetic susceptibility.

The static susceptibility of uranium compounds is mostly determined by
the localised magnetic moments concentrated on the uranium ions   \cite{Troc2012,Rogalev2015,Sanchez2017,Khmelevsky2022}. 
For each particular $x,y,z$   direction along the $ a,b,c$ orthorhombic crystallographic axis
(say along $b$-axis) it can be written as \cite{Mineev2017} 
\begin{equation}
\chi_{yy}({\bf k})=\chi_b({\bf k})= \frac{1}{\chi^{-1}_b(T,H_b)+2\gamma^b_{lm}k_lk_m},
\label{b}
\end{equation}
where $\chi_b(T,H_b)$ is the temperature and field dependent homogeneous part of static susceptibility.
The imaginary part of frequency dependent susceptibility according to  Kramers and Kronig  is related  with  static susceptibility 
\begin{equation}
\chi_b({\bf k})=\frac{1}{\pi}\fint_{-\infty}^{+\infty}\frac{\chi_b^{\prime\prime}({\bf k},\omega)}{\omega}d\omega.
\label{KK}
\end{equation}
The simplest form of frequency dependent susceptibility satisfying the Kramers-Kronig relationship  is
\begin{equation}
\chi_b({\bf k},\omega)=\frac{1}{-i\omega\tau_b+\chi_b^{-1}({\bf k})}
\label{b1}
\end{equation}
so that 
its imaginary part is
\begin{equation}
\chi^{\prime\prime}_b({\bf k},\omega)=\frac{\omega\tau_b}{(\omega\tau_b)^2+\chi_b^{-2}({\bf k})}.
\label{Imm}
\end{equation}

The contribution of conducting electrons does not exceed 10 percent of the total magnetisation ( see the Ref.29,30). Under these conditions, it is reasonable to consider  a conducting electron gas moving in an anisotropic continuum ferromagnetic matrix and interacting according to Eq.(1).
A change in the magnetic susceptibility of itinerant electrons due to interaction (1) is insignificant compared to the field and temperature dependence in magnetic fluctuations determined by the continuum ferromagnetic matrix.

In URhGe and UCoGe, the band structure and spectrum of electronic excitations are not well known and it is pointless to take into account the inter-band spin-orbit coupling between undefined bands, as well as the anisotropy of the g-factor of conducting electrons. Thus, to study our problem about the field dependence of the effective mass of the electron, it is sufficient  to limit ourselves to one conduction band, split by the exchange and external fields. The very formation of the heavy fermion band through the Kondo effect occurs at temperatures much higher than the temperatures at which the ferromagnetic superconducting state exists. Therefore, this is beyond the scope of our consideration.

The matrix of electron Green function \cite{Book}  is written using  the  normal $G_{\alpha\beta}({\bf k},i\omega_n)$ and the  Gor'kov Green function $ F_{\alpha\beta}({\bf k},i\omega_n)$
\begin{equation}
{\bf G}_{\alpha\beta}({\bf k},i\omega_n)=\left(\begin{array}{cc}G_{\alpha\beta}({\bf k},i\omega_n)&-F_{\alpha\beta}({\bf k},i\omega_n)\\
-F\dagger_{\alpha\beta}({\bf k},i\omega_n)&-G^t_{\alpha\beta}(-{\bf k},-i\omega_n)\end{array}\right).
\end{equation}
It is determined by the Dyson-Eliashberg  equation 
\begin{equation}
{\bf G}^{-1}_{\alpha\beta}({\bf k},i\omega_n)=\left(\begin{array}{cc}i\omega_n\delta_{\alpha\beta}-H_{\alpha\beta}({\bf k})-\Sigma_{\alpha\beta}({\bf k},i\omega_n)&-\Phi_{\alpha\beta}({\bf k},i\omega_n)\\
-\Phi^\dagger_{\alpha\beta}({\bf k},i\omega_n)&i\omega_n\delta_{\alpha\beta}+H^t_{\alpha\beta}({\bf k})+\Sigma^t_{\alpha\beta}({\bf k},-i\omega_n)\end{array}\right).
\end{equation}
Here, $\omega_n=\pi(2n+1)$ are the fermion Matsubara frequencies, the superscript $"t"$ implies transposition. The one particle energy
\begin{equation}
H_{\alpha\beta}({\bf k})=\xi_{\bf k}\delta_{\alpha\beta}-\mu_{ij}({\bf k})\sigma^i_{\alpha\beta}(h_j+H_j)
\end{equation}
consists of kinetic
\begin{equation}
\xi_{\bf k}=\varepsilon_{\bf k}-\mu
\end{equation}
and the Zeeman energy including the electron spin  interaction with internal field ${\bf h}$ produced by spontaneous magnetisation and the external field 
${\bf H}$. $\sigma^i_{\alpha\beta}=(\sigma^x_{\alpha\beta},\sigma^y_{\alpha\beta},\sigma^z_{\alpha\beta})$  are the  Pauli matrices in the spin space. In what follows we will put $\mu_{ij}({\bf k})=\mu_B\delta_{ij}$ ignoring  tensor character and wave vector dependence of the Zeeman interaction induced by spin-orbit coupling.

The "normal"  part of  self-energy matrix
\begin{equation}
\Sigma_{\alpha\beta}({\bf k},i\omega_n)=[\delta_{\alpha\beta}-Z_{\alpha\beta}({\bf k},i\omega_n)]i\omega_n
\end{equation}
is determined from the self-consistency equation
\begin{equation}
\Sigma_{\alpha\beta}({\bf k},i\omega_n)
=g^2T\sum_{\omega_m}\int\frac{d^3{\bf k}^\prime}{(2\pi)^3}\sigma^i_{\alpha\lambda}
\chi_{ij}({\bf k}-{\bf k}^\prime, i\omega_n-i\omega_m)G_{\lambda\gamma}({\bf k}^\prime,i\omega_m)\sigma_{\gamma\beta}^j.
\label{sigma}
\end{equation}

Effective mass  $m$ of an electron in metal differs from the bare electron mass due to static and dynamic electron-phonon interaction with crystal lattice. We will be interested in
the additional contribution to electron effective mass arising 
due to electron-electron interaction through the spin fluctuations exchange. 
Following to the treatment  described in review \cite{Scalapino1969}  
let us use the spectral representations for the electron Green function 
\begin{equation}
G_{\alpha\beta}({\bf k},i\omega_n)=-\int_{-\infty}^{\infty}\frac{d\omega^\prime}{\pi}\frac{Im~G_{\alpha\beta}({\bf k},\omega^\prime+i\delta)}{i\omega_n-\omega^\prime}
\label{Gr}
\end{equation}
and for the boson propagator
\begin{equation}
\chi_{ij}({\bf k}, i\omega_\nu)=-\int_0^\infty \frac{d\Omega}{\pi}\chi^{\prime\prime}_{ij}({\bf k},\Omega)\left (\frac{1}{i\omega_\nu-\Omega}-\frac{1}{i\omega_\nu+\Omega } \right ),
\label{Chi}
\end{equation}
where $\chi^{\prime\prime}_{ij}({\bf k},\Omega)$ is the imaginary part of $\chi_{ij}({\bf k},\Omega)$.
Substituting these expressions to Eq.(\ref{sigma}) and performing the analytical continuation \cite{Scalapino1969}  we come  to
\begin{eqnarray}
\Sigma_{\alpha\beta}({\bf k},\omega)=g^2\int_{-\infty}^\infty\frac{d\omega^\prime}{\pi}\int_0^\infty \frac{d\Omega}{\pi} \int\frac{d^3{\bf k}^\prime}{(2\pi)^3}
\sigma^i_{\alpha\lambda}\chi^{\prime\prime}_{ij}({\bf k}-{\bf k}^\prime, \Omega)Im~G_{\lambda\gamma}({\bf k}^\prime,\omega^\prime+i\delta)
\sigma_{\gamma\beta}^j\nonumber\\
\times\left [\frac{f(-\omega^\prime)+n(\Omega)}{\omega^\prime+\Omega-\omega}+\frac{f(\omega^\prime)+n(\Omega)}{\omega^\prime-\Omega-\omega}\right],~~~~~~~~~~~~~
\label{sigma1}
\end{eqnarray}
where $f(\omega)=(\exp(\omega/T)+1)^{-1}$ and $n(\Omega)=(\exp(\Omega/T)-1)^{-1}$ are the Fermi and the Bose distribution functions correspondingly.

Writing the self-energy matrix in the form
\begin{equation}
\Sigma_{\alpha\beta}({\bf k},i\omega_n)=\Sigma({\bf k},i\omega_n)\delta_{\alpha\beta}+{\bf \Sigma}({\bf k},i\omega_n)\mbox{\boldmath $\sigma$}_{\alpha\beta},
\end{equation}
we obtain the Green function  
\begin{eqnarray}
G_{\lambda\gamma}({\bf k}^\prime,\omega^\prime+i\delta)=
\frac{1}{2}\left\{ \left(\delta_{\lambda\gamma} +\frac{\mu_B({\bf h}+{\bf H})-{\bf \Sigma}}{|\mu_B({\bf h}+{\bf H}-{\bf \Sigma}|}\mbox{\boldmath $\sigma$}_{\lambda\gamma}\right)\left(
\omega^\prime+i\delta-\Sigma-
\xi_{{\bf k}^\prime}+|\mu_B({\bf h}+{\bf H}-{\bf \Sigma}|\right)^{-1}
 \right.\nonumber\\
\left.
 +\left(\delta_{\lambda\gamma} -\frac{\mu_B({\bf h}+{\bf H})-{\bf \Sigma}}{|\mu_B({\bf h}+{\bf H}-{\bf \Sigma}|}\mbox{\boldmath $\sigma$}_{\lambda\gamma}\right)\left(
\omega^\prime+i\delta-\Sigma-
\xi_{{\bf k}^\prime}-|\mu_B({\bf h}+{\bf H}-{\bf \Sigma}|\right)^{-1}\right\}.
\label{G}
\end{eqnarray}
Integration over $\xi_{{\bf k}^\prime}$ in Eq.(\ref{sigma1}) fixes the value of ${\bf k}^\prime$ on the Fermi surfaces determined by the equations
\begin{equation}
\varepsilon_{\bf k}-\mu=\pm |\mu_B({\bf h}+{\bf H})|
\label{FS}
\end{equation}
corresponding to two terms in Eq.(\ref{G}). 
Performing integration over $\xi_{{\bf k}^\prime}$ in Eq.(\ref{sigma}) 
we obtain for the scalar and vector parts of the self-energy the following equations:
\begin{eqnarray}
\Sigma({\bf k},\omega)=-\frac{g^2}{2}\int_{-\infty}^\infty d\omega^\prime\int_0^\infty \frac{d\Omega}{\pi}~ 
\left [ \langle N_{0+}({\bf k}^\prime)\chi^{\prime\prime}_{ii}({\bf k}-{\bf k}^\prime,\Omega)\rangle_++\langle N_{0-}({\bf k}^\prime)
\chi^{\prime\prime}_{ii}({\bf k}-{\bf k}^\prime,\Omega)\rangle_-\right]\nonumber\\
\left [\frac{f(-\omega^\prime)+n(\Omega)}{\omega^\prime+\Omega-\omega}+\frac{f(\omega^\prime)+n(\Omega)}{\omega^\prime-\Omega-\omega}\right],
\label{scalar}
\end{eqnarray}
\begin{eqnarray}
{\bf \Sigma}_j({\bf k},\omega)
=-\frac{g^2}{2}\int_{-\infty}^\infty d\omega^\prime\int_0^\infty \frac{d\Omega}{\pi}~ 
\left [\langle N_{0+}({\bf k}^\prime)\Phi_j({\bf k}-{\bf k}^\prime,\Omega)\rangle_+-\langle N_{0-}({\bf k}^\prime)\Phi_j({\bf k}-{\bf k}^\prime,\Omega)
\rangle_-\right ]\nonumber\\
\left [\frac{f(-\omega^\prime)+n(\Omega)}{\omega^\prime+\Omega-\omega}+\frac{f(\omega^\prime)+n(\Omega)}{\omega^\prime-\Omega-\omega}\right],~
\label{vector}
\end{eqnarray} 
where $\langle...\rangle_+ $ and $\langle...\rangle_+ $ mean the angle averaging over the Fermi surfaces determined by Eq.(\ref{FS}) with the  density of states
$N_{0+}({\bf k})$ and $N_{0-}({\bf k})$.
 Deriving Eqs.(\ref{scalar}),(\ref{vector}) we used the symmetry of  the susceptibility tensor $\chi_{ij}=\chi_{ji}$, so that
 $
 \sigma^i_{\alpha\gamma}\chi^{\prime\prime}_{ij}\sigma^j_{\gamma\beta}=\chi^{\prime\prime}_{ii}\delta_{\alpha\beta}
  $
 and
 \begin{equation}
 \chi^{\prime\prime}_{ij}\sigma^i_{\alpha\lambda}\sigma^k_{\lambda\gamma}\sigma^j_{\gamma\beta}\hat\nu_k=\left[2\chi^{\prime\prime}_{ij}\hat\nu_i-\chi^{\prime\prime}_{ii}\hat\nu_j\right ]\sigma^j_{\alpha\beta}=\Phi_j\sigma^j_{\alpha\beta}.
 \label{21}
 \end{equation}
 We are looking for the expression ${\bf \Sigma}_j({\bf k},\omega))$ linear in frequency. In this case, it is sufficient to use the components of the unit vector 
 $\hat\nu_j=
(\mu_B({\bf h}+{\bf H})-{\bf \Sigma})_j/|\mu_B({\bf h}+{\bf H})-{\bf \Sigma}|$ at $\omega=0$:
\begin{equation}
\hat\nu_j=\frac{({\bf h}+{\bf H})_j}{|{\bf h}+{\bf H}|}.
\label{nu}
\end{equation}

The frequency dependent terms in denominators  in two terms of the Green function Eq.(\ref{G}) are
\begin{equation}
\omega-\Sigma\mp\hat\nu_j{\bf \Sigma}_j=\omega-(1-Z)\omega\pm\hat\nu_jZ_j\omega.
\end{equation}
Hence, the effective masses on two Fermi surfaces Eq.(\ref{FS}) 
\begin{equation}
m^\star_\pm({\bf k})=m[Z({\bf k})\pm\hat\nu_jZ_j({\bf k})]
\label{m}
\end{equation}
are the functions of momentum  on the corresponding Fermi surface.

Let us determine $Z$ and $Z_j$.
The Eqs. (\ref{scalar}) and (\ref{vector}) can be rewritten as
\begin{eqnarray}
\Sigma({\bf k},\omega)=(1-Z({\bf k},\omega))\omega=\frac{g^2}{2}\int_0^\infty d\omega^\prime\int_0^\infty \frac{d\Omega}{\pi}~ 
\left [ \langle N_{0+}({\bf k}^\prime)\chi^{\prime\prime}_{ii}({\bf k}-{\bf k}^\prime,\Omega)\rangle_++\langle N_{0-}({\bf k}^\prime)
\chi^{\prime\prime}_{ii}({\bf k}-{\bf k}^\prime,\Omega)\rangle_-\right]
\nonumber\\
\times\left[(f(-\omega^{\prime})+n(\Omega))\left (  \frac{1}{\omega^\prime+\Omega+\omega}-  \frac{1}{\omega^\prime+\Omega-\omega}\right)\right.~~~
\nonumber\\
+\left.
(f(\omega^{\prime})+n(\Omega))\left (  \frac{1}{-\omega^\prime+\Omega+\omega}-  \frac{1}{-\omega^\prime+\Omega-\omega}\right)\right],~~~
\end{eqnarray}
\begin{eqnarray}
\Sigma_j({\bf k},\omega)=-Z_j({\bf k},\omega))\omega=\frac{g^2}{2}
\int_{0}^\infty d\omega^\prime \int_0^\infty \frac{d\Omega}{\pi}~ 
\left [\langle N_{0+}({\bf k}^\prime)\Phi_j({\bf k}-{\bf k}^\prime,\Omega)\rangle_+-\langle N_{0-}({\bf k}^\prime)\Phi_j({\bf k}-{\bf k}^\prime,\Omega)
\rangle_-\right ]
\nonumber\\
\times \left[(f(-\omega^{\prime})+n(\Omega))\left (  \frac{1}{\omega^\prime+\Omega+\omega}-  \frac{1}{\omega^\prime+\Omega-\omega}\right)\right.\nonumber\\+\left.
(f(\omega^{\prime})+n(\Omega))\left (  \frac{1}{-\omega^\prime+\Omega+\omega}-  \frac{1}{-\omega^\prime+\Omega-\omega}\right)\right].~~~
\end{eqnarray}
Similar to the calculations in model with electron-phonon interaction \cite{Scalapino1969,McMillan1968} in low temperature limit 
we put $n(\Omega)=0$. Writing then $f(-\omega^{\prime})=1-f(\omega^{\prime})$  and performing integration over $\omega^\prime$     we come at $\omega=0$ and $T=0$ to
\begin{equation}
1-Z({\bf k})=-\frac{g^2}{\pi}\int_0^\infty \frac{d\Omega}{\Omega}\left [ \langle N_{0+}({\bf k}^\prime)\chi^{\prime\prime}_{ii}({\bf k}-{\bf k}^\prime,\Omega)\rangle_++\langle N_{0-}({\bf k}^\prime)
\chi^{\prime\prime}_{ii}({\bf k}-{\bf k}^\prime,\Omega)\rangle_-\right],
\label{sc}
\end{equation}
\begin{equation}
Z_j({\bf k})=\frac{g^2}{\pi}\int_0^\infty \frac{d\Omega}{\Omega}
\left [\langle N_{0+}({\bf k}^\prime)\Phi_j({\bf k}-{\bf k}^\prime,\Omega)\rangle_+-\langle N_{0-}({\bf k}^\prime)\Phi_j({\bf k}-{\bf k}^\prime,\Omega)
\rangle_-\right ].
\label{ve}
\end{equation}
Remembering the Kramers-Kronig relation (\ref{KK}) we obtain
\begin{equation}
1-Z({\bf k})=-\frac{g^2}{2}\left [ \langle N_{0+}({\bf k}^\prime)\chi_{ii}({\bf k}-{\bf k}^\prime)\rangle_++\langle N_{0-}({\bf k}^\prime)
\chi_{ii}({\bf k}-{\bf k}^\prime)\rangle_-\right],
\label{sca}
\end{equation}
\begin{equation}
Z_j({\bf k})=\frac{g^2}{2}
\left [\langle N_{0+}({\bf k}^\prime)\Phi_j({\bf k}-{\bf k}^\prime)\rangle_+-\langle N_{0-}({\bf k}^\prime)\Phi_j({\bf k}-{\bf k}^\prime)
\rangle_-\right ].
\label{vec}
\end{equation}

Thus, the effective mass is expressed through  the angle averages over the Fermi surfaces from various   magnetic susceptibility components $
\chi_{ij}({\bf k})= (\chi^{-1}_{ij}(T,H)+2\gamma^{ij}_{lm}k_lk_m)^{-1}$ where $\chi_{ij}(T,H)$ are functions of magnetic field and temperature determined experimentally.
Structure of Fermi surfaces in URhGe and UCoGe is unknown.
Moreover,
in the absence of a microscopic theory, the momentum dependence of susceptibilities 
is also unknown. 
The indeterminacy can be removed by means of the simple assumption.
Namely, in what follows,  we will assume that the inhomogeneous part of the inverse susceptibility $\gamma^{ij}_{lm}k_lk_m$
 is much less than its homogeneous part $\chi^{-1}_{ij}(T,H_b )$. Note, that the low temperature static susceptibility $\chi_{ij}(T,H)$ in uranium compounds along various crystallographic axes is of the order  $10^{-3}$ per unit volume \cite{Hardi2011}. 

The neglect of momentum dependence of magnetic fluctuations allows rewrite Eqs.(\ref{sca}) and (\ref{vec}) in more simple form
\begin{equation}
1-Z=-\frac{N_{0+}+N_{0-}}{2}g^2\chi_{ii}({\bf H},T)),
\end{equation}
\begin{equation}
Z_j=\frac{N_{0+}-N_{0-}}{2}g^2
\left[2\chi_{ij}({\bf H},T))\hat\nu_i-\chi_{ii}({\bf H},T))\hat\nu_j\right ],
\end{equation}
where $ N_{0+}=\langle N_{0+}({\bf k}\rangle_ +$ and   $ N_{0-}=\langle N_{0-}({\bf k}\rangle_ -$ are the average values of density of state on the corresponding Fermi surface.

Thus, the effective mass is 
\begin{equation}
m^\star_\pm=m(1+\lambda_\pm)
\end{equation}
and
\begin{equation}
\lambda_\pm=\frac{g^2}{2}\left\{(N_{0+}+N_{0-})\chi_{ii}
\pm
(N_{0+}-N_{0-})
\left[2\chi_{ij}\hat\nu_i\hat\nu_j-\chi_{ii}\right]\right\},
\end{equation}
such that the full effect of effective mass renormalisation is
\begin{equation}
\lambda_++\lambda_-=g^2(N_{0+}+N_{0-})\chi_{ii}({\bf H},T).
\label{34}
\end{equation}

\section{Superconducting critical temperature \\in neglect of orbital effects}

 Temperature of transition to superconducting state is determined from the linearised equation for the superconducting self-energy
$\Phi_{\alpha\beta}({\bf k},i\omega_n)$. 
The equation for the superconducting part of self-energy is obtained after transformation of interaction (\ref{int}) to the sum of two terms corresponding to singlet and triplet pairing (see  the derivation in the paper \cite{Sam}). 
We are interested in the superconducting state with triplet pairing. The singlet part can be neglected  in view of paramagnetic depairing leading to the lowering of transition temperature. Thus, we have
\begin{equation}
\Phi_{\alpha\beta}({\bf k},i\omega_n)=-g^2T\sum_{\omega_m}\int\frac{d^3{\bf k}^\prime}{(2\pi)^3}(i\sigma^i\sigma^y)_{\beta\alpha}
W_{ij}({\bf k},{\bf k}^\prime, i\omega_n-i\omega_m)
(i\sigma^j\sigma^y)^\dagger_{\lambda\mu}F_{\lambda\mu}({\bf k}^\prime,i\omega_m),
\label{Phi}
\end{equation}
where 
\begin{equation}
W_{ij}({\bf k},{\bf k}^\prime, i\omega_n-i\omega_m)
=-\left (\frac{1}{2}\chi^u_{ll}({\bf k},{\bf k}^\prime, i\omega_n-i\omega_m)\delta_{ij}-\chi^u_{ij}({\bf k},{\bf k}^\prime, i\omega_n-i\omega_m)\right).
\end {equation}

Here, 
\begin{equation}
\chi^u_{ij}({\bf k},{\bf k}^\prime, i\omega_\nu)=\frac{1}{2}\left (\chi_{ij}({\bf k}-{\bf k}^\prime,i\omega_\nu)-\chi_{ij}({\bf k}+{\bf k}^\prime,i\omega_\nu)\right)
\end{equation}
is the part of susceptibility odd in respect of both  arguments ${\bf k}$ and ${\bf k}^\prime$.
In this chapter we will work  with susceptibility in the diagonal form, that is assuming $\chi_{ij}=0$ at $i\ne j$.  The odd part of the susceptibility
 extracted 
from Eqs.(\ref{b}) and (\ref{b1}) is:
\begin{equation}
\chi^u_{ij}({\bf k},{\bf k}^\prime, i\omega_\nu)\approx\frac{4\gamma^{ij}_{lm}k_l k_m^\prime}{(\omega_\nu\tau_{ij}+\chi_{ij}^{-1}(T,H))^2}
= 4\gamma^{ij}_{lm}k_lk^\prime_m\chi^2_{ij}(i\omega_\nu).
\label{D}
\end{equation}
Here, after explicitly identifying the angular dependence, we have neglected  by the momentum dependence in  $\chi(i\omega_\nu)$, as it was done in the effective mass derivation.
To avoid confusion, let us point out  that this expression does not contain summation over repeating indices $ij$.

The spectral representations for the Gor'kov Green function and the odd part of susceptibility have the same form as corresponding "normal" spectral representations given by Eqs.(\ref{Gr}),(\ref{Chi})
\begin{equation}
F_{\lambda\mu}({\bf k},i\omega_n)=-\int_{-\infty}^{\infty}\frac{d\omega^\prime}{\pi}\frac{Im~F_{\lambda\mu}({\bf k},\omega^\prime+i\delta)}{i\omega_n-\omega^\prime},
\label{F}
\end{equation}
\begin{equation}
\chi^u_{ij}(\hat{\bf k},\hat{\bf k}^\prime, i\omega_\nu)=-\int_0^\infty \frac{d\Omega}{\pi}~ \chi^{u\prime\prime }_{ij}(\hat{\bf k},\hat{\bf k}^\prime,\Omega)\left (\frac{1}{i\omega_\nu-\Omega}-\frac{1}{i\omega_\nu+\Omega } \right ).
\label{Chi^u}
\end{equation}

 Substituting these expressions to Eq.(\ref{Phi}) and performing the analytical continuation we come  to
\begin{eqnarray}
\Phi_{\alpha\beta}(\hat{\bf k},\omega)=-g^2\int_{-\infty}^\infty\frac{d\omega^\prime}{\pi}\int_0^\infty \frac{d\Omega}{\pi} \int\frac{d^3{\bf k}^\prime}{(2\pi)^3}
(i\sigma^i\sigma^y)_{\alpha\beta}
Im~W_{ij}(\hat{\bf k},\hat{\bf k}^\prime, \Omega)(i\sigma^i\sigma^y)^\dagger_{\lambda\mu}Im~F_{\lambda\mu}({\bf k}^\prime,\omega^\prime+i\delta)\nonumber
\\
\times\left [\frac{f(-\omega^\prime)+n(\Omega)}{\omega^\prime+\Omega-\omega}+\frac{f(\omega^\prime)+n(\Omega)}{\omega^\prime-\Omega-\omega}\right],~~~~~~~~~~~~~
\label{Phi2}
\end{eqnarray}
where the Gor'kov Green function in linear in respect to  $\Phi_{\gamma\delta}({\bf k}^\prime,\omega^\prime)$ approximation is expressed through the product of the normal Green functions
\begin{equation}
F_{\lambda\mu}({\bf k}^\prime,\omega^\prime+i\delta)=G_{\lambda\gamma}({\bf k}^\prime,\omega^\prime+i\delta)\Phi_{\gamma\delta}({\bf k}^\prime,\omega^\prime)G_{\mu\delta}(-{\bf k}^\prime,-(\omega^\prime+i\delta)).
\end{equation}
In the absence of field or when the external magnetic field ${\bf H}$ is parallel to the spontaneous magnetisation ${\bf h}$
the normal Green function  is the diagonal matrix \begin{equation}
G_{\lambda\gamma}=\left(\begin{array}{cc} G^\uparrow&0\\ 0&G^\downarrow \end{array} \right),
\end{equation}
where
\begin{equation}
G^{\uparrow,\downarrow}({\bf k}^\prime,\omega^\prime+i\delta)=\frac{1}{\omega^\prime+i\delta-\xi_{{\bf k}^\prime}  \pm\mu_B(h+H)-\Sigma\mp\Sigma_z }
\end{equation}
corresponds to electrons in conduction band split by the exchange and external field on two bands with spin-up and spin-down.
The situation with external field directed perpendicular to spontaneous magnetisation we will discuss at the end of this  chapter.

The matrix for the self-energy of superconducting state with triplet pairing is
\begin{equation}
\Phi_{\alpha\beta}=\left(\begin{array}{cc} \Phi^\uparrow&\Phi^0\\ \Phi^0&\Phi^\downarrow   \end{array} \right),
\end{equation}
According to Eq.(\ref{Phi2})  its components  satisfy to the system of linear integral equations, which can be written symbolically as
\begin{eqnarray}
\Phi^l=\hat A^{lm}\Phi^m,
\end{eqnarray}
where $\Phi^l=(\Phi^\uparrow,\Phi^\downarrow,\Phi^0)$ and $\hat A^{lm}$ is the matrix of integral operators. In the case of diagonal matrix 
 of susceptibility 
the equations for $\Phi^\uparrow$ and $\Phi^\downarrow$ components of self-energy split  from the  the equation for the $\Phi^0$. This is easy to check performing multiplication of matrices in Eq.(\ref{Phi2}). We will consider only the system
for $\Phi^\uparrow$ and $\Phi^\downarrow$ which corresponds to the so called equal-spin pairing superconducting state.
After performing all integrations the system is
transformed to the system of algebraic equations.
The critical temperature is determined by the equality of determinant of this system to zero. 
 
 The equations for $\Phi^\uparrow$ and $\Phi^\downarrow$ are
 \begin{eqnarray}
\Phi^\uparrow(\hat{\bf k},\omega)=g^2\int_{-\infty}^\infty\frac{d\omega^\prime}{\pi}\int_0^\infty \frac{d\Omega}{\pi} \int\frac{d^3{\bf k}^\prime}{(2\pi)^3}\nonumber\\
\times\left\{
\chi^{u\prime\prime}_{zz}(\hat{\bf k},\hat{\bf k}^\prime, \Omega)
Im\left[G^\uparrow({\bf k}^\prime,\omega^\prime+i\delta)\Phi^\uparrow(\hat{\bf k}^\prime,\omega^\prime)G^\uparrow(-{\bf k}^\prime,-(\omega^\prime+i\delta))\right]\right.~~~\nonumber
\\
\left.+(\chi^{u\prime\prime}_{xx}(\hat{\bf k},\hat{\bf k}^\prime, \Omega)-\chi^{u\prime\prime}_{yy}(\hat{\bf k},\hat{\bf k}^\prime, \Omega))
Im\left[G^\downarrow({\bf k}^\prime,\omega^\prime+i\delta)\Phi^\downarrow(\hat{\bf k}^\prime,\omega^\prime)G^\downarrow(-{\bf k}^\prime,-(\omega^\prime+i\delta))\right]\right\}\nonumber\\
\left [\frac{f(-\omega^\prime)+n(\Omega)}{\omega^\prime+\Omega-\omega}+\frac{f(\omega^\prime)+n(\Omega)}{\omega^\prime-\Omega-\omega}\right],~~
\end{eqnarray}
 \begin{eqnarray}
\Phi^\downarrow(\hat{\bf k},\omega)=g^2\int_{-\infty}^\infty\frac{d\omega^\prime}{\pi}\int_0^\infty \frac{d\Omega}{\pi} \int\frac{d^3{\bf k}^\prime}{(2\pi)^3}\nonumber\\
\times\left\{
(\chi^{u\prime\prime}_{xx}(\hat{\bf k},\hat{\bf k}^\prime, \Omega)-\chi^{u\prime\prime}_{yy}(\hat{\bf k},\hat{\bf k}^\prime, \Omega))
Im\left[G^\uparrow({\bf k}^\prime,\omega^\prime+i\delta)\Phi^\uparrow(\hat{\bf k}^\prime,\omega^\prime)G^\uparrow(-{\bf k}^\prime,-(\omega^\prime+i\delta))\right]
\right.\nonumber
\\
+\left.\chi^{u\prime\prime}_{zz}(\hat{\bf k},\hat{\bf k}^\prime, \Omega)
Im\left[G^\downarrow({\bf k}^\prime,\omega^\prime+i\delta)\Phi^\downarrow(\hat{\bf k}^\prime,\omega^\prime)G^\downarrow(-{\bf k}^\prime,-(\omega^\prime+i\delta))\right]\right\}\nonumber\\
\left [\frac{f(-\omega^\prime)+n(\Omega)}{\omega^\prime+\Omega-\omega}+\frac{f(\omega^\prime)+n(\Omega)}{\omega^\prime-\Omega-\omega}\right].~~~~~~~~~
\end{eqnarray}

The integration over momenta is the integration over energy and over the Fermi surface for the spin-up and spin-down electron band
\begin{equation}
\int\frac{d^3{\bf k}^\prime}{(2\pi)^3}=\int d\xi_{{\bf k}^\prime}\int \frac{dS_{\hat{\bf k}^\prime}}{v_F^\prime}N_{0\pm}(\hat{\bf k}^\prime).
\end{equation}
Performing integration over $\xi_{{\bf k}^\prime}$ 
we obtain for 
$$
\Phi^\uparrow(\hat{\bf k},\omega)=(Z+\hat\nu_jZ_j)\Delta^\uparrow(\hat{\bf k},\omega),~~~~~~~~~~~
\Phi^\downarrow(\hat{\bf k},\omega)=(Z-\hat\nu_jZ_j)\Delta^\downarrow(\hat{\bf k},\omega)
$$
the following expressions
\begin{eqnarray}
(Z+\hat\nu_jZ_j)\Delta^\uparrow(\hat{\bf k},\omega)=g^2\int_{-\infty}^\infty\frac{d\omega^\prime}{\omega^\prime}\int_0^\infty \frac{d\Omega}{\pi}
\int \frac{dS_{\hat{\bf k}^\prime}}{v_F^\prime}~~~~~~~~~~~~~~~~~~~~~~~~~\nonumber\\
\times\left [N_{0+}(\hat{\bf k}^\prime)\chi^{u\prime\prime}_{zz}(\hat{\bf k},\hat{\bf k}^\prime, \Omega)\Delta^\uparrow(\hat{\bf k}^\prime,\omega^\prime)+
N_{0-}(\hat{\bf k}^\prime)(\chi^{u\prime\prime}_{xx}(\hat{\bf k},\hat{\bf k}^\prime, \Omega)-\chi^{u\prime\prime}_{yy}(\hat{\bf k},\hat{\bf k}^\prime, \Omega))\Delta^\downarrow(\hat{\bf k}^\prime,\omega^\prime)\right]\nonumber\\
\times\left [\frac{f(-\omega^\prime)+N(\Omega)}{\omega^\prime+\Omega-\omega}+\frac{f(\omega^\prime)+N(\Omega)}{\omega^\prime-\Omega-\omega}\right],
\label{Eq1}
\end{eqnarray}
\begin{eqnarray}
(Z-\hat\nu_jZ_j)\Delta^\downarrow(\hat{\bf k},\omega)=g^2\int_{-\infty}^\infty\frac{d\omega^\prime}{\omega^\prime}\int_0^\infty \frac{d\Omega}{\pi}
\int \frac{dS_{\hat{\bf k}^\prime}}{v_F^\prime}~~~~~~~~~~~~~~~~~~~~~~~~~\nonumber\\
\times \left [N_{0+}(\hat{\bf k}^\prime)
(\chi^{u\prime\prime}_{xx}(\hat{\bf k},\hat{\bf k}^\prime, \Omega)-\chi^{u\prime\prime}_{yy}(\hat{\bf k},\hat{\bf k}^\prime, \Omega))\Delta^\uparrow(\hat{\bf k}^\prime,\omega^\prime)+N_{0-}(\hat{\bf k}^\prime)\chi^{u\prime\prime}_{zz}(\hat{\bf k},\hat{\bf k}^\prime, \Omega)
\Delta^\downarrow (\hat{\bf k}^\prime,\omega^\prime)\right]\nonumber\\
\times
\left [\frac{f(-\omega^\prime)+N(\Omega)}{\omega^\prime+\Omega-\omega}+\frac{f(\omega^\prime)+N(\Omega)}{\omega^\prime-\Omega-\omega}\right].
\label{Eq2}
\end{eqnarray}
We  will consider $B$-superconducting state \cite{Mineev2017} with the order parameter 
\begin{eqnarray}
\Delta^\uparrow(\hat{\bf k},\omega)=\eta^\uparrow(\omega)\hat k_z,\nonumber\\
\Delta^\downarrow(\hat{\bf k},\omega)=\eta^\downarrow(\omega)\hat k_z.
\label{OP}
\end{eqnarray}
The treatment of $A$-state is much more cumbersome.

Substituting equation (\ref{D}) in Eqs.(\ref{Eq1}),  (\ref{Eq2})  we obtain
\begin{eqnarray}
(Z+\hat\nu_jZ_j)\eta^\uparrow(\omega)=g^2\int_{-\infty}^\infty\frac{d\omega^\prime}{\omega^\prime}\int_0^\infty \frac{d\Omega}{\pi}\nonumber\\
\times\left\{A_+\gamma_{zz}^{zz}[\chi^2_{zz}( \Omega)]^{{\prime\prime}}\eta^\uparrow(\omega^\prime)+
A_-(\gamma_{zz}^{xx}[\chi^2_{xx}( \Omega)]^{\prime\prime}-\gamma_{zz}^{yy}[\chi^{2}_{yy}( \Omega)]^{\prime\prime})\eta^\downarrow(\omega^\prime)\right\}
\nonumber\\
\times\left [\frac{f(-\omega^\prime)+n(\Omega)}{\omega^\prime+\Omega-\omega}+\frac{f(\omega^\prime)+n(\Omega)}{\omega^\prime-\Omega-\omega}\right],~~~~~~~~~~~~~~~
\label{Eq3}
\end{eqnarray}
\begin{eqnarray}
(Z-\hat\nu_jZ_j)\eta^\uparrow(\omega)=g^2\int_{-\infty}^\infty\frac{d\omega^\prime}{\omega^\prime}\int_0^\infty \frac{d\Omega}{\pi}\nonumber\\
\times
\left\{A_+(\gamma_{zz}^{xx}[\chi^{2}_{xx}( \Omega)]^{\prime\prime}-\gamma_{zz}^{yy}[\chi^{2}_{yy}( \Omega)]^{\prime\prime})\eta^\uparrow(\omega^\prime)+
A_-\gamma_{zz}^{zz}[\chi^{2}_{zz}( \Omega)]^{\prime\prime}\eta^\downarrow(\omega^\prime)\right\}
\nonumber\\
\times\left [\frac{f(-\omega^\prime)+n(\Omega)}{\omega^\prime+\Omega-\omega}+\frac{f(\omega^\prime)+n(\Omega)}{\omega^\prime-\Omega-\omega}\right],~~~~~~~~~~~~~~~
\label{Eq4}
\end{eqnarray}
where 
\begin{equation}
A_\pm=4\langle  k_z^2N_{0\pm}(\hat{\bf k})\rangle,
\end{equation} 
and $\langle...\rangle$ means averaging over the Fermi surface.

These equations can be rewritten as
\begin{eqnarray}
(Z+\hat\nu_jZ_j)\eta^\uparrow(\omega)=g^2\int_0^\infty\frac{d\omega^\prime}{\omega^\prime}\int_0^\infty \frac{d\Omega}{\pi}\nonumber\\
\times\left\{A_+\gamma_{zz}^{zz}[\chi^{2}_{zz}( \Omega)]^{\prime\prime}\eta^\uparrow(\omega^\prime)+
A_-(\gamma_{zz}^{xx}[\chi^{2}_{xx}( \Omega)]^{\prime\prime}-\gamma_{zz}^{yy}[\chi^{2}_{yy}( \Omega)]^{\prime\prime})\eta^\downarrow(\omega^\prime)\right\}
\nonumber\\
\times
\left[
(f(-\omega^\prime)+n(\Omega))\left (\frac{1}{\omega^\prime+\Omega-\omega}+\frac{1}
{\omega^\prime+\Omega+\omega}
\right)\right.\nonumber\\
\left.
-(f(\omega^\prime)+n(\Omega))   \left(\frac{1}
{-\omega^\prime+\Omega-\omega}+\frac{1}
{-\omega^\prime+\Omega+\omega}\right)\right],
\label{Eq5}
\end{eqnarray}
\begin{eqnarray}
(Z-\hat\nu_jZ_j)\eta^\downarrow(\omega)=g^2\int_0^\infty\frac{d\omega^\prime}{\omega^\prime}\int_0^\infty \frac{d\Omega}{\pi}\nonumber\\
\times
\left\{A_+(\gamma_{zz}^{xx}[\chi^{2}_{xx}( \Omega)]^{\prime\prime}-\gamma_{zz}^{yy}[\chi^{2}_{yy}( \Omega)]^{\prime\prime})\eta^\uparrow(\omega^\prime)+
A_-\gamma_{zz}^{zz}[\chi^{2}_{zz}( \Omega)]^{\prime\prime}\eta^\downarrow(\omega^\prime)\right\}
\nonumber\\
\times 
\left[
(f(-\omega^\prime)+n(\Omega))\left (\frac{1}{\omega^\prime+\Omega-\omega}+\frac{1}
{\omega^\prime+\Omega+\omega}
\right)\right.\nonumber\\
\left.
-(f(\omega^\prime)+n(\Omega))   \left(\frac{1}
{-\omega^\prime+\Omega-\omega}+\frac{1}
{-\omega^\prime+\Omega+\omega}\right)\right]
\label{Eq6}
\end{eqnarray}

Similar to the calculations in model with electron-phonon interaction \cite{Scalapino1969,McMillan1968} in low temperature limit 
one must  put $n(\Omega)=0$ and also introduce the cut-off  in the integral over $\omega^\prime$. 
The cut-off $\omega_0=(\tau\chi(T,H))^{-1}$ is determined by the limit of validity of quasi-elastic form of frequency dependent susceptibility
$\chi_c(\omega)=(-i\omega\tau+\chi_c^{-1})^{-1}$. Its value far enough from the Curie temperature is of  the order  several Kelvin \cite{Huxley2003,Huxley2011}  but tends to zero at $T\to T_{Curie}$.
 Introducing a cut-off $\omega_0$ in the integral over $\omega^\prime$ we get in low frequency limit
\begin{eqnarray}
(1+\lambda_+)\eta^\uparrow\cong\left [\lambda^\uparrow\eta^\uparrow+\lambda^{\uparrow\downarrow}\eta^\uparrow\right]\ln\frac{\omega_0}{T_c},\\
(1+\lambda_-)\eta^\downarrow\cong\left[\lambda^{\downarrow\uparrow}\eta^\uparrow+\lambda^\downarrow\eta^\downarrow\right]\ln\frac{\omega_0}{T_c},
\end{eqnarray}
where was used $(Z\pm\hat\nu_jZ_j)=1+\lambda_\pm$ and
\begin{eqnarray}
\lambda^\uparrow=g^2A_+\frac{2}{\pi}\int_0^\infty\frac{d\Omega}{\Omega}\gamma_{zz}^{zz}[\chi^{2}_{zz}( \Omega)]^{\prime\prime},~~~~~~~~~\\
\lambda^{\uparrow\downarrow}=g^2A_-\frac{2}{\pi}\int_0^\infty\frac{d\Omega}{\Omega}
\left(\gamma_{zz}^{xx}[\chi^{2}_{xx}( \Omega)]^{\prime\prime}-\gamma_{zz}^{yy}[\chi^{2}_{yy}( \Omega)]^{\prime\prime}\right),\\
\lambda^{\downarrow\uparrow}=g^2A_+\frac{2}{\pi}\int_0^\infty\frac{d\Omega}{\Omega}
\left(\gamma_{zz}^{xx}[\chi^{2}_{xx}( \Omega)]^{\prime\prime}-\gamma_{zz}^{yy}[\chi^{2}_{yy}( \Omega)]^{\prime\prime}\right),\\
\lambda^\downarrow=g^2A_-\frac{2}{\pi}\int_0^\infty\frac{d\Omega}{\Omega}\gamma_{zz}^{zz}[\chi^{2}_{zz}( \Omega)]^{\prime\prime}.~~~~~~~~~
\end{eqnarray}
Performing  integration over $\Omega$ we obtain
\begin{eqnarray}
\lambda^\uparrow=g^2A_+\gamma_{zz}^{zz}\chi^{2}_{zz}( T,H),~~~~~~~~~~\\
\lambda^{\uparrow\downarrow}=g^2A_-
\left(\gamma_{zz}^{xx}\chi^{2}_{xx}(T,H)-\gamma_{zz}^{yy}\chi^{2}_{yy}(T,H)\right),\\
\lambda^{\downarrow\uparrow}=g^2A_+
\left(\gamma_{zz}^{xx}\chi^{2}_{xx}(T,H)-\gamma_{zz}^{yy}\chi^{2}_{yy}(T,H)\right),\\
\lambda^\downarrow=g^2A_-\gamma_{zz}^{zz}\chi^{2}_{zz}(T,H).~~~~~~~~~~
\end{eqnarray}
Equating the determinant of system (58), (59) we come to the formula for the critical temperature in neglect of orbital effects
\begin{equation}
T_c=\omega_0\exp \left (  -\frac{1}{\Lambda} \right),
\end{equation}
where the constant of interaction 
\begin{equation}
\Lambda=\frac{1}{2}\left( \frac{\lambda^\uparrow}{1+\lambda_+}+ \frac{\lambda^\downarrow}{1+\lambda_-} \right)+
\sqrt{ \frac{1}{4}\left( \frac{\lambda^\uparrow}{1+\lambda_+}- \frac{\lambda^\downarrow}{1+\lambda_-} \right)^2+\frac{\lambda^{\uparrow\downarrow}\lambda^{\downarrow\uparrow}}{(1+\lambda_+)(1+\lambda_-)}  }
\label{Lambda}
\end{equation}
is expressed through  the static susceptibility components according to equations (33) and  (64)-(67).

Now, we  present the modification of obtained results in the situation when  the external magnetic field ${\bf H}=H\hat y$ is directed perpendicular to spontaneous magnetisation ${\bf h}=h\hat z$. In this case the initial point symmetry group of orthorhombic ferromagnet $G=U(1)\times(E, C_2^z,RC_2^x,RC_2^y)$ decreases to monoclinic one $G=U(1)\times(E, RC_2^x)$. Here $U(1)$ is the group of gauge transformation and $R$ is the time reversal operation.
It is natural to choose the spin quantisation axis along the direction $\mbox{\boldmath${\nu}$} $ of the total magnetic field $h\hat z+H\hat y$  given by  Eq.(\ref{nu}). The multiplications of spin matrices in Eq.(\ref{Phi2}) have to be performed in this basis. The order parameter (\ref{OP}) corresponding to equal spin pairing superconducting state with spin parallel and antiparallel to direction  $\mbox{\boldmath${\nu}$} $ is still approprate by symmetry.
Repeating the calculations similar to those were described above we come to formula for the critical temperature
\begin{equation}
T_c(\varphi)=\omega_0\exp \left (  -\frac{1}{\Lambda_\varphi}\right),
\end{equation}
where formula for $\Lambda_\varphi$ is obtained from Eq.(\ref{Lambda})
 by means the following modifications
\begin{eqnarray}
\gamma_{zz}^{zz}\chi_{zz}^2\to  \gamma_{zz}^{zz}\chi_{zz}^2\cos^2\varphi+ \gamma_{zz}^{yy}\chi_{yy}^2\sin^2\varphi,\\
\gamma_{zz}^{yy}\chi_{yy}^2\to  \gamma_{zz}^{zz}\chi_{zz}^2\sin^2\varphi+ \gamma_{zz}^{yy}\chi_{yy}^2\cos^2\varphi
\end{eqnarray}
in all expressions for  $\lambda_+,\lambda_-,\lambda^\uparrow, \lambda^\downarrow,\lambda^{\uparrow\downarrow}, \lambda^{\downarrow\uparrow}$.
  Here, the angle $\varphi$ is defined by $\tan\varphi=H/h$.

\section{Discussion}

\subsection{URhGe}

The renormalisation of the effective mass (\ref{34}) is proportional to the trace of the magnetic susceptibility. Crystallographic direction $a$ is the magnetically hard axis, and the susceptibility in this direction can be neglected, therefore in magnetic field parallel to $b$-axis
\begin{equation}
\lambda_++\lambda_-\cong g^2(N_{0+}+N_{0-})\left [\chi_b(H_b)+\chi_c(H_b)\right ].
\label{lambda1}
\end{equation}
The susceptibility along the $b$ axis is strongly enhanced near the metamagnetic transition, which was established by direct measurements\cite{Nakamura2017}.
On the other hand, this result can be understood at using the NMR relaxation rate found in the paper \cite{Tokunaga2015}.
Indeed, the transverse relaxation rate $1/T_2$ serves as a measure
 of  intensity of magnetic oscillations along the direction of the field.
   It is expressed through the imaginary part of the magnetic susceptibility $\chi^{\prime\prime}_b({\bf k},\omega)$ in the low-frequency limit $\omega\ll\omega_{{\text {NMR}} }$ and through hyperfine coupling constant ${\cal A}_b({\bf k})$
\begin{equation}
\frac{1}{T_{2,b}}
\propto T\int\frac{d^3{\bf k}}{(2\pi)^3}|{\cal A}_b({\bf k})|^2
\frac{\chi^{\prime\prime}_b({\bf k},\omega)}{\omega}.
\end{equation}
Here we have neglected the so-called Redfield contribution
\cite{Tokunaga2015}, which is expressed through the longitudinal relaxation rate $1/T_{1,b}$. The latter
less than $1/T_{2,b}$
more than an order of magnitude.
Using the equation (\ref{Imm}), we can evaluate the integral 
\begin{equation}
\frac{1}{T_{2,b}}
\propto
  T\int\frac{d^3{\bf k}}{(2\pi)^3}|{\cal A}_b({\bf k})|^2\frac{\tau_b}{(\omega\tau_b)^2+\chi_b^{-2}({\bf k} )}.
   \label{rateloc}
\end{equation}
The wave vectors  in the denominator do not exceed the inverse interatomic distance, therefore, at temperatures not too close to $T_{Curie}$, neglecting in the integrand by the dependence on the wave vector
we obtain in the limit of low frequencies
\begin{equation}
\frac{1}{T_{2,b}}
   \propto nT|{\cal A}_b|^2\chi^2_b(T,H_b)\tau_b,
   \label{rateloc}
\end{equation}
where $n$ is the volume of an elementary cell of reciprocal space.
We see that the increase in $1/T_{2,b}$ reported in \cite{Tokunaga2015} corresponds to an increase in static susceptibility.

Susceptibility in the $c$ direction $\chi_c(H_b)$ depending on the magnetic field in the $b$ direction
not measured. However, as noted in \cite{Mineev2017,Mineev2021} $\chi_c(H_b)$ increases with decreasing Curie temperature $T_{Curie}(H_b)$ . On the other hand, this fact is in agreement with NMR  relaxation rate $1/T_{1,b}$ measured in  \cite{Tokunaga2015}.
Indeed, this relaxation rate is expressed through magnetisation fluctuations in directions perpendicular to the field direction.
In URhGe $1/T_{1,b}$ is determined primarily by magnetic fluctuations along the $c$-crystallographic direction
\begin{equation}
\frac{1}{T_{1,b}}
\propto T\int\frac{d^3{\bf k}}{(2\pi)^3}|{\cal A}_c({\bf k})|^2
\frac{\chi^{\prime\prime}_c({\bf k},\omega)}{\omega},
\end{equation}
probed at frequency $\omega=\omega_{NMR}$.
Being less than $1/T_{2,b}$, the longitudinal relaxation rate  $1/T_{1,b}$
   also tends to increase \cite{Tokunaga2015}.
So, at the assumption $\omega_{NMR}\tau_c\ll\chi_c^{-1}$, similar to (\ref{rateloc}),
we arrive to
\begin{equation}
\frac{1}{T_{1,b}}
\propto
   \int\frac{d^3{\bf k}}{(2\pi)^3}\frac{\tau_c}{(\omega\tau_c)^2+\chi_c^{ -2}({\bf k})}
   \propto nT|{\cal A}_c|^2\chi_c^2(T,H_b)\tau_c.
   \label{rate1}
\end{equation}
We see that the increase in $1/T_{1,b}$ is associated with an increase in the static susceptibility $\chi_c(H_b)$.

Thus, both $\chi_b(H_b)$ and $\chi_c(H_b)$ increase as we approach the metamagnetic transition,   resulting in an increase in the effective mass according to Eq.
(\ref{lambda1}).

\subsection{UCoGe}

In UCoGe, the susceptibility $\chi_b(H_b)$ remains constant up to fields above 40 T~ \cite{Knafo2012}. Therefore, the observed increase in effective mass \cite{Aoki2014} is associated with
increase in susceptibility $\chi_c(H_b)$. Indeed, it increases due to the suppression of the Curie temperature, which is confirmed by the analysis of NMR data presented in the article \cite{Ishida2021}, where the stimulation of superconductivity when approaching the ferromagnetic critical region was also qualitatively explained.

In weak fields parallel to the $c$ axis, the magnetic susceptibility $\chi_c(H_c)$ gradually decreases \cite{Huy2008}, which is accompanied by a decrease in the effective mass and suppression of the pairing amplitude \cite{Wu2017} .

\subsection{UTe$_2$}

In UTe$_2$ the situation is more complicated.
NMR measurements in a field along the $b$-crystallographic axis by Y.Tokunaga et al. \cite{Tokunaga2023} demonstrate a strong increase in the intensity of longitudinal magnetic fluctuations in fields above 15 Tesla.
    However, in contrast to URhGe, the $1/T_{2,b}$ enhancement at $H_b>$ 15 T is not associated with an increase in static susceptibility
    $\chi_b(H_b)$. The latter remains constant almost until the metamagnetic transition at 34.5 T (see articles \cite{Miyake2019,Miyake2021}). Consequently, an increase in the NMR relaxation rate $1/T_{2,b}$
    at least not too close to the metamagnetic transition, does not correspond to the naive estimation given by the equation ({\ref{rateloc}).
  
    On the other hand, growth
    the rate of longitudinal relaxation $1/T_{1,b}$ with $H_b$, which is also reported in \cite{Tokunaga2023}, may occur due to the enhancement
    susceptibility $\chi_a(H_b)$ and $ \chi_c(H_b)$ in the directions $a$ and $c$ as a function of the field $H_b$. In this material, $a$ is the easy magnetic axis.
    Thus, the assumption about the strengthening of $\chi_a(H_b)$ explains both the growth of
    longitudinal relaxation rate $1/T_{1,b}$ and 
an increase in the effective mass according to $\lambda\cong N_0g^2\chi_a(H_b)$. The increase of effective mass with field $H_b$  was established experimentally in the works \cite{Imajo2019,Miyake2019,Miyake2021}.

The single-band approach described in the previous chapter can explain the stimulation of superconductivity near the metamagnetic transition, but certainly not in the field range $(15~T<H_b<25~T)$.
We will have to leave this problem for future research.

\section {Orbital effects}

Let us consider qualitatively the field dependence of critical temperature of transition to superconducting state including the orbital effects.
To avoid cumbersome expressions  we limit ourselves by the single band spin-up superconducting state.
In the case of single band spin-up superconducting state the critical temperature in neglect of orbital effects is
\begin{equation}
T_c=\omega_0\exp \left (  -\frac{1+\lambda_+}{\lambda^\uparrow} \right),
\label{sb}
\end{equation}
which is formally similar to the classic McMillan formula.
The parameters determining the critical temperature are
\begin{equation}
\lambda_+(\varphi)=g^2N_{0+}(\chi_{zz}\cos^2\varphi+ \chi_{yy}\sin^2\varphi),
\end{equation}
\begin{equation}
\lambda^{\uparrow}(\varphi)=4g^2\langle k_z^2N_{0+}({\bf k})\rangle(\gamma_{zz}^{zz}\chi_{zz}^2\cos^2\varphi+ \gamma_{zz}^{yy}\chi_{yy}^2\sin^2\varphi).
\end{equation}
Both of these values increase following increase of susceptibility components as we get closer to metamagnetic transition in URhGe. The growth of $\lambda^{\uparrow}$ proportional to square of susceptibility components is faster than the growth of $\lambda_+$. Thus, the critical temperature in neglect of orbital depairing effectively increases in vicinity of metamagnetic transition.

The obtained  field dependence of effective mass and  $T_c$  determined by the magnetic field depending susceptibility allows make conclusion about unusual behaviour of critical temperature of transition to the superconducting state taking into account orbital effects.
Indeed, the temperature dependence of the upper critical field has a standard form, parametrically depending on $T_c$ and the Fermi velocity. In particular,  the upper critical field  at zero temperature $H_{c2}(T=0)=H_0$ is 
\begin{equation}
H_0=c\Phi_0\frac{T_c^2}{\hbar ^2v_F^2},
\label{H_0}
\end{equation}
where $\Phi_0=\pi\hbar c/e$ is the magnetic flux quantum and $c$ is a numerical constant. 
 The upper critical field near the zero temperature is
 \begin{equation}
 H=H_0\left(1-d\frac{T^2}{T_c^2}\right),
 \end{equation}
 where $d$ is numerical constant of the order of unity. Thus, in low temperature-high field
region the actual critical temperature is
 \begin{equation}
 T_c^{orb}=\frac{\hbar v_F}{\sqrt{cd\Phi_0}}\sqrt{H_0-H}.
 \label{T_c}
 \end{equation}
  The finite critical temperature is realised at $H_0>H$. In small fields $H\leq H_0$ along $b$-axis, where the $T_c$ and $v_F$ are determined by their zero-field values, Eq.(\ref{T_c}) presents the usual  dependence of the critical
temperature from magnetic field.  It follows by the fields interval where $H>H_0$ and superconducting state is absent. 
At higher fields in the vicinity of the metamagnetic transition, $T_c(H)$ and $1/v_F(H)\propto (1+\lambda_+)$ increase rapidly. Accordingly, $H_0(H)$ has a sharp maximum such that $H_0(H)>H$. As a result, a solution to the equation (\ref{T_c}) arises again, corresponding to
   the appearance of a reentrant superconducting state.

\section{Conclusion}

A theory of superconductivity in ferromagnetic uranium compounds URhGe and UCoGe has been developed, based on electron-electron interaction through magnetic fluctuations, providing a natural explanation for the magnetic field dependence of the effective mass of electrons and the intensity of pairing interaction.
The resulting formulas establish a connection between the field dependence of two independently measured physical quantities: the effective mass and the magnetic susceptibility components, which is in reasonable qualitative agreement with existing experimental data.

The established field-dependent growth  of  the Fermi velocity and the   critical temperature in neglect of orbital effects presents mechanism 
of reentrance of superconducting state in vicinity of metamagnetic transition in URhGe and sharp increase of the upper critical field in UCoGe at the Curie temperature decreasing.

   As for UTe$_2$, the theory provides a plausible explanation for the increase in effective mass in a field parallel to the $b$ axis, but does not explain
   the appearance of a reentrant superconducting state at field $H_b \approx15~T$ which is quite far from the metamagnetic transition.

\end{document}